\documentclass[prl,twocolumn,preprintnumbers,superscriptaddress,showpacs]{revtex4}
\usepackage{amssymb}
\usepackage{graphicx}
\usepackage{dcolumn}
\usepackage{bm}
\usepackage{multirow}

\begin{document}

\title{Room-temperature multiferroic hexagonal LuFeO$_3$ films}

\author{Wenbin Wang}
\affiliation{Department of Physics, University of Tennessee, Knoxville, TN 37996, USA}
\affiliation{Materials Science and Technology Division, Oak Ridge National Laboratory, Oak Ridge, TN 37831, USA}

\author{Jun Zhao}
\affiliation{Department of Physics, University of California, Berkeley, CA 94720, USA}

\author{Wenbo Wang}
\affiliation{Department of Physics and Astronomy, Rutgers University, Piscataway, NJ 08854, USA}

\author{Zheng Gai}
\affiliation{Center for Nanophase Materials Sciences, Oak Ridge National Laboratory, Oak Ridge, TN 37831, USA}

\author{Nina Balke}
\affiliation{Materials Science Division, Argonne National Laboratory, Argonne, IL 60439, USA}

\author{Miaofang Chi}
\affiliation{Materials Science and Technology Division, Oak Ridge National Laboratory, Oak Ridge, TN 37831, USA}

\author{Ho Nyung Lee}
\affiliation{Materials Science and Technology Division, Oak Ridge National Laboratory, Oak Ridge, TN 37831, USA}

\author{Wei Tian}
\affiliation{Quantum Condensed Matter Division, Oak Ridge National Laboratory, Oak Ridge, TN 37831, USA}

\author{Leyi Zhu}
\affiliation{Materials Science Division, Argonne National Laboratory, Argonne, IL 60439, USA}

\author{Xuemei Cheng}
\affiliation{Department of Physics, Bryn Mawr College, Bryn Mawr, PA 19010, USA}

\author{David J. Keavney}
\affiliation{Advanced Photon Source, Argonne National Laboratory, Argonne, IL 60439, USA}

\author{Jieyu Yi}
\affiliation{Department of Physics, University of Tennessee, Knoxville, TN 37996, USA}

\author{Thomas Z. Ward}
\affiliation{Materials Science and Technology Division, Oak Ridge National Laboratory, Oak Ridge, TN 37831, USA}

\author{Paul C. Snijders}
\affiliation{Materials Science and Technology Division, Oak Ridge National Laboratory, Oak Ridge, TN 37831, USA}

\author{Hans M. Christen}
\affiliation{Materials Science and Technology Division, Oak Ridge National Laboratory, Oak Ridge, TN 37831, USA}

\author{Weida Wu}
\affiliation{Department of Physics and Astronomy, Rutgers University, Piscataway, NJ 08854, USA}

\author{Jian Shen$^*$}
\affiliation{Department of Physics, University of Tennessee, Knoxville, TN 37996, USA}
\affiliation{State Key Laboratory of Surface Physics and Department of Physics, Fudan University, Shanghai 200433, China}

\author{Xiaoshan Xu$^*$}
\affiliation{Materials Science and Technology Division, Oak Ridge National Laboratory, Oak Ridge, TN 37831, USA}

\date{\today}

\begin{abstract}
 The crystal and magnetic structures of single-crystalline hexagonal LuFeO$_3$ films have been studied using x-ray, electron and neutron diffraction methods.
 The polar structure of these films are found to persist up to 1050 K; and the switchability of the polar behavior is observed at room temperature, indicating ferroelectricity.
 An antiferromagnetic order was shown to occur below 440 K, followed by a spin reorientation resulting in a weak ferromagnetic order below 130 K.
 This observation of coexisting multiple ferroic orders demonstrates that hexagonal LuFeO$_3$ films are room-temperature multiferroics.
\end{abstract}


\maketitle

\clearpage
 The coexistence and coupling of ferroelectric and magnetic orders in multiferroic materials promise many improvements over singly ordered ferroic materials for next generation applications in information technology, sensing, and actuation.\cite{Schmid1994,Khomskii2009,Spaldin2010} 
 For widespread implementation of this technology, coexistence of long range magnetic and electric orders at room temperature will be required; at present, there is only one material, BiFeO$_3$, known to exhibit ferroelectricity and antiferromagnetic order above room temperature.\cite{Wang2003} 
 Hexagonal ferrites (h-RFeO$_3$, R=Sc, Y, Ho-Lu) are expected to be ferroelectric due to the polar structure that lies at the origin of the ferroelectricity of YMnO$_3$ above room temperature ($T_C$$\approx$1000 K).\cite{Fennie2005}
 In fact, the evidence of ferroelectricity has been found in h-YbFeO$_3$ films below 470 K.\cite{Jeong2012}
 Antiferromagnetic spin structures involving a triangular arrangement of the moments in the $a$-$b$ plane are expected in h-RFeO$_3$ due to the structural symmetry.\cite{Munoz2000}
 Despite the frustration created by the triangular lattice, the strong interactions between the Fe$^{3+}$ sites due to high spin and large Fe-O interactions \cite{Wang2012,Cheremisinoff1990} are expected to greatly increase the magnetic ordering temperature of h-RFeO$_3$ compared with that of RMnO$_3$ ($T_N$$\approx$100 K)\cite{Fiebig2000}.
 This makes h-RFeO$_3$ promising candidates to be room-temperature multiferroics. 
 
 Here we focus on LuFeO$_3$, which is known to crystallize in both orthorhombic (o-LuFeO$_3$) and presently investigated hexagonal (h-LuFeO$_3$) structures.
 Orthorhombic LuFeO$_3$ is stable in bulk with a non-polar Pbnm structure and exhibits C-type antiferromagnetism below $T_N$=620 K.\cite{Cheremisinoff1990}
 In contrast, a polar P6$_3$cm structure has been found in bulk (metastable) and thin film h-LuFeO$_3$ at room temperature.\cite{Bossak2004,Magome2010}
 A weak ferromagnetic order below 120 K has been recently reported in h-LuFeO$_3$ films.\cite{Akbashev2011}
 While the polar structure of h-LuFeO$_3$ suggests ferroelectricity, the structural phase transitions have not been investigated. 
 More importantly, the magnetic ordering temperature (the aforementioned key factor of the room-temperature multiferroicity in h-RFeO$_3$) and the detailed magnetic structure have not been unambiguously determined in h-LuFeO$_3$.
 In this work, we show using a combined investigation of neutron scattering and magnetometry that an antiferromagnetic order occurs in h-LuFeO$_3$ films below $T_N$=440 K, followed by a spin reorientation below $T_R$=130 K as the origin of the previously observed weak ferromagnetic order. 
 We also show that the polar structure P6$_3$cm in h-LuFeO$_3$ films persists up to 1050 K according to the electron and x-ray diffraction measurements.
 The switchability of the polar behavior is observed using piezoelectric force microscopy at room-temperature, providing strong evidence of ferroelectricity.
 These observations demonstrate that the h-LuFeO$_3$ films are multiferroic at room temperature.
 
 Hexagonal LuFeO$_3$ films (20-60 nm) have been grown epitaxially using pulsed laser deposition both onto $\alpha$-Al$_2$O$_3$ (001) substrates and onto sputtered epitaxial Pt films (30 nm) on $\alpha$-Al$_2$O$_3$ (001) substrates.\cite{NoteStrain} 
 The film structure was determined \textit{in-situ} by reflection high energy electron diffraction (RHEED) and \textit{ex-situ} by x-ray diffraction (XRD) and transmission electron microscopy (TEM) data.\cite{NoteRHEED}
 Piezoresponse force microscopy (PFM) was studied on the h-LuFeO$_3$/Pt/Al$_2$O$_3$ films.
 The neutron diffraction experiments were carried out on a stacked h-LuFeO$_3$ films on the HB1A thermal triple axis spectrometer at the High Flux Isotope Reactor, Oak Ridge National Laboratory.
 The temperature dependence of the magnetization was measured using a superconducting quantum interference device (SQUID) magnetometer.
 The temperature dependence of x-ray spectroscopy was studied using polarized synchrotron at beam line 4-ID-C at the Advanced Photon Source.

\begin{figure}[b]
\centerline{
\includegraphics[width = 3.5 in]{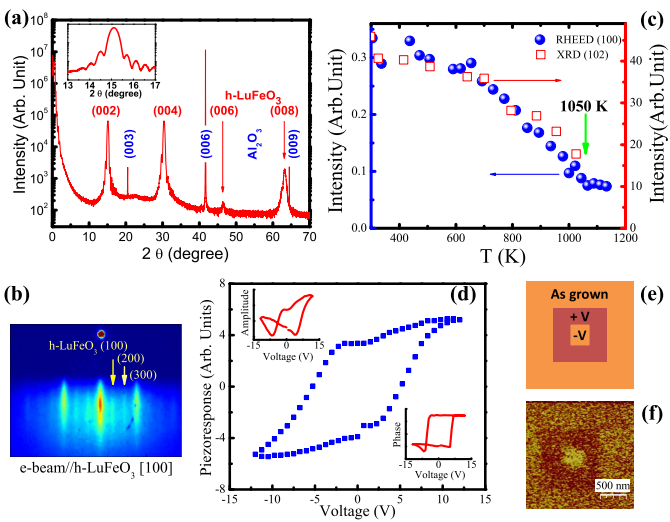}}
\caption{ (Color online)
 The structural characterizations of the h-LuFeO$_3$ films using RHEED and XRD. 
 (a) A typical $\theta$-2$\theta$ scan of an h-LuFeO$_3$ film. 
 The inset shows the close-up view of the h-LuFeO$_3$ (002) peak.
 (b) A RHEED image of the h-LuFeO$_3$ film with electron beam along Al$_2$O$_3$ and h-LuFeO$_3$ [100] direction. 
 (c) Intensities of RHEED (100) peak and XRD (102) peak as functions of temperature. 
 (d) PFM response displaying square-shaped hysteresis loop. 
 The amplitude and phase are shown in the insets.
 (e) Schematic of written domain pattern with DC voltage ({\it V} = 20 V$_{dc}$).
 (f) PFM image of the same region without DC voltage.
}
\label{fig_structFE}
\end{figure}

\begin{figure}[b]
\centerline{
\includegraphics[width = 3.5 in]{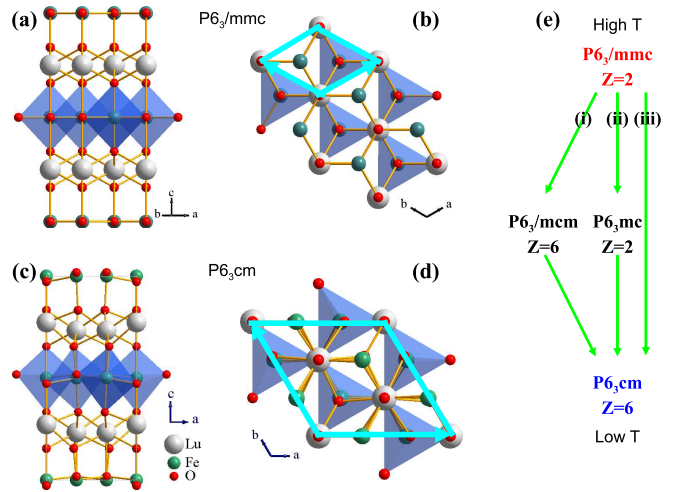}}
\caption{ (Color online)
 The schematic representation of P6$_3$/mmc and P6$_3$cm structures and the possible transition routes.
 (a) and (b) are the P6$_3$/mmc structure viewed from [120] and [001] direction respectively.
 (c) and (d) are the P6$_3$cm structure viewed from [100] and [001] direction respectively.
  The bold lines in (b) and (d) indicate the unit cells in the $a$-$b$ plane.
 (e) Possible routes from P6$_3$/mmc to P6$_3$cm structures.
}
\label{fig_struct}
\end{figure}

 As shown in Fig. \ref{fig_structFE}(a), the x-ray diffraction data indicate that the h-LuFeO$_3$ films are grown along the [001] direction without detectable impurity phases.
 The RHEED image shows intense diffraction streaks separated by weak streaks (Fig. \ref{fig_structFE}(b)), which can be understood in terms of the detailed structure of h-LuFeO$_3$.
 The P6$_3$cm structure can be viewed as a result of the distortions from a non-polar P6$_3$/mmc structure with alternating LuO$_2$ and FeO layers (Fig. \ref{fig_struct} (a)).
 The distortions include the rotations of the FeO$_5$ trigonal bipyramids with propagation vector $\vec{K}$=(1/3,1/3,0) corresponding to a $\sqrt{3}$$\times$$\sqrt{3}$ reconstruction in the $a$-$b$ plane (tripling the unit cell) and the displacements of the Lu atoms along $c$-axis with $\vec{K}$=(0,0,0) which breaks the inversion symmetry and allows for a electric polarization along the $c$-axis (Fig. \ref{fig_struct} (a-d)).\cite{Fennie2005}
 The RHEED pattern is in perfect agreement with the $\sqrt{3}$$\times$$\sqrt{3}$ reconstruction in the $a$-$b$ plane.
 From the streak separations, the in-plane lattice constants of the h-LuFeO$_3$ films can be estimated as $a$=5.96 +/- 0.1 \AA, consistent with the value of the bulk P6$_3$cm structure.\cite{Magome2010}
 Hence the indices of the diffraction streaks can be assigned using a P6$_3$cm unit cell,\cite{NoteUnitCell} as indicated in Fig. \ref{fig_structFE}(b), suggesting an eptitaxial relation Al$_2$O$_3$ [100]//h-LuFeO$_3$ [100], in agreement with the XRD characterization.\cite{supplementary}

 Temperature dependent RHEED and XRD have been carried out to study the robustness of the polar structure.
 As shown in Fig. \ref{fig_structFE}(c), the intensity of the RHEED (100) peak (normalized with that of the (300) peak) is displayed as a function of temperature. 
 A clear transition at approximately 1050 (+/-50) K is observed from the RHEED (100) peak intensity, indicating a structural transition. 
 The intensity of the XRD (102) peak follows the trend of the RHEED (100) peak closely, although the transition is not observed due to limited temperature range.\cite{Note100}

 Symmetry analysis predicts three possible routes for the transition from a P6$_3$cm (Z=6, polar) structure to a P6$_3$/mmc (Z=2, non-polar) structure, where Z is the number of formula unit per unit cell (Fig. \ref{fig_struct}(e)).\cite{Fennie2005,Lonkai2004}
 According to the kinetic theory, the diffraction intensities differ dramatically for (100) and (102) peaks for different structures:\cite{supplementary}
 Firstly, both (100) and (102) peaks vanish for structures with Z=2 unit cells. 
 Secondly, for a P6$_3$/mcm structure, the diffraction intensity for (100) and (102) peaks should be similar, while for a P6$_3$cm structure, the (100) peak should be much weaker than that of the (102) peak.
 Therefore, the diminishing intensities of the RHEED (100) peak and XRD (102) peak suggest a transition from a Z=6 to a Z=2 structure.
 The fact that the intensity of the XRD (102) peak follows that of the RHEED (100) peak indicates the structure of the h-LuFeO$_3$ films at $T$$<$1050 K belongs to a P6$_3$cm structure, i.e. no intermediate P6$_3$/mcm structure involved.    Hence the RHEED and XRD data suggest that the polar structure P6$_3$cm of the h-LuFeO$_3$ films persists up to at least 1050 K.

 To assess possible ferroelectricity in our h-LuFeO$_3$ films given the polar structure, we studied the switchability of the h-LuFeO$_3$ films (30 nm) on Pt-buffered Al$_2$O$_3$ using PFM.\cite{Lines1977} 
 Here we used Band Excitation Switching Spectroscopy with an ac imaging voltage of 2 V$_{ac}$ and a maximum switching voltage of 12 V$_{dc}$ to measure the piezoresponse loops.\cite{Maksymovych2008}
 A metal coated tip (Nanosensors) is used. 
 We measured three piezoresponse loops at each position of a 10$\times$10 point grid in a 3$\times$3 $\mu$m$^2$ area and extracted the average switching behavior as displayed in Fig. \ref{fig_structFE}(d). 
 The piezoelectric loops show square-shaped loops with switching voltages around 5 V$_{dc}$ and -7 V$_{dc}$, which are indicative of ferroelectric switching.
 Further evidence of ferroelectric switching was obtained in decoupled write/read PFM processes, i.e. poling (write) domain pattern with DC voltage (+/- 20 V$_{dc}$) on the h-LuFeO3 film as shown in Fig. \ref{fig_structFE}(e), then visualizing (read) the same region with only AC imaging voltage. 
 As shown in Fig. \ref{fig_structFE}(f), the written domain patterns is clearly visible in the PFM image. 
 Furthermore, the pattern persist at least two hours without any visible decay in domain contrast.\cite{supplementary}
  
\begin{figure}[b]
\centerline{
\includegraphics[width = 3.3 in]{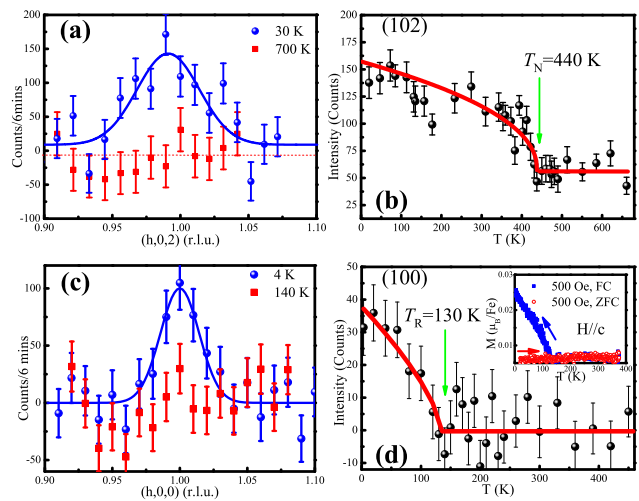}}
\caption{ (Color online)
 Magnetic Bragg peaks and the temperature dependence of their intensities from neutron diffraction. 
 (a) and (c) are the diffraction intensity profile near the (102) and (100) magnetic Bragg points respectively. 
 The lines are the Gaussian fits. 
 (b) and (d) are the temperature dependence of the (102) and (100) magnetic Bragg peak intensities repectively. 
 The curves in (b) and (d) are fits to the data points (see text). 
 The inset of (d) shows the temperature dependence of the magnetization in ZFC and FC processes with the magnetic field along the $c$-axis.
}
\label{fig_TN}
\end{figure}

\begin{figure}[b]
\centerline{
\includegraphics[width = 3.5 in]{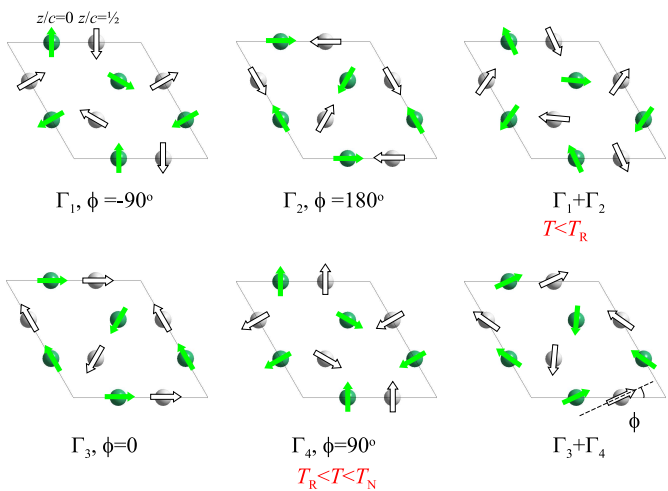}}
\caption{ (Color online)
 The possible magnetic structures of Fe$^{3+}$ sites in P6$_3$cm crystallographic structure and the assignments of magnetic structures for h-LuFeO$_3$. The darker (lighter) balls represent Fe$^{3+}$ sites at z/c=0 (z/c=1/2) and the arrows indicate the magnetic moment orientation.  
 $\Gamma_1$ ($\Gamma_3$) and $\Gamma_2$ ($\Gamma_4$) are related by the in-plane rotation angle $\phi$. 
 The $\Gamma_1$ ($\Gamma_2$) and $\Gamma_3$ ($\Gamma_4$) are a homometric pair, meaning indistinguishable in our neutron scattering measurement. 
 The magnetic structure of h-LuFeO$_3$ takes the form $\Gamma_2$, mixed with $\Gamma_1$ at $T$$<T_R$ and $\Gamma_4$ at $T_R$$<$$T$$<$$T_N$.
}
\label{fig_magstructure}
\end{figure}

 Now that the strong evidence of the room-temperature ferroelectricity in h-LuFeO$_3$ films is obtained, we investigate how robust the magnetic ordering is.
 Neutron diffraction experiments were carried out to study the magnetic order of h-LuFeO$_3$ films. 
 The films were aligned in the (h 0 l) diffraction plane in a closed-cycle refrigerator. 
 At low temperature, several Bragg peaks were clearly observed. 
 As shown in Fig. \ref{fig_TN} (a) and (b), the (102) peak intensity drops as the temperature increases up to approximately 440 K while for the (100) peak the transition occurs at approximately 130 K (Fig. \ref{fig_TN} (c) and (d)). 
 In contrast, the change of the diffraction peaks (300) and (004) with temperature is minimal between 4 K and 450 K.\cite{supplementary} 
 In addition to neutron diffraction, measurements using SQUID magnetometer revealed that the zero-field-cool (ZFC) magnetization and field-cool (FC) magnetization split at 130 K under magnetic field along the $c$-axis, indicating the appearance of a ferromagnetic component along the $c$-axis (Fig. \ref{fig_TN} (d) inset).

 Since h-LuFeO$_3$ is isomorphic with RMnO$_3$ and their magnetic structures can be characterized with the same propagation vector $\vec{K}$=(0,0,0) (meaning magnetic unit cell is the same as the structural unit cell), we can analyze the magnetic structure of h-LuFeO$_3$ following the symmetry analysis in RMnO$_3$.\cite{Bertaut1963,Bertaut1968,Munoz2000,Brown2006} 
 In this case, the Bragg peaks (102) and (100) are in principle coming from a combination of nuclear and magnetic diffractions. 
 However, our structural characterizations do not indicate any clear transition close to 440 K (Fig. \ref{fig_structFE} (c)), which is consistent with the fact that in RMnO$_3$ the magnetic diffraction constitutes the majority of the intensity for (102) peak.\cite{Munoz2000} 
 Therefore, the transition at 440 K corresponds to the appearance of a long range antiferromagnetic order. 
 Taking $T_N$=440 (+/-10) K, one can fit between 200 K$<$$T$$<$440 K the peak intensity using $I=I_n+I_m(1-T/T_N)^\alpha$ where $I_n$ and $I_m$ are the nuclear and magnetic contributions of the intensity respectively. 
 The result shows $\alpha$=0.46 +/- 0.08, falling well in the reported value range of YMnO$_3$.\cite{Chatterji2007,Roessli2005} 
 The fit for (100) peak intensity gives a larger value $\alpha$ = 0.69 +/- 0.15. 
 The development of a ferromagnetic component below $T_R$=130 (+/-1) K suggests that the system displays a second magnetic phase by a spin reorientation which is consistent with our x-ray magnetic dichroism measurements\cite{supplementary}.

 The possible magnetic structures of the h-LuFeO$_3$ films can be assigned by taking both the neutron diffraction and the magnetization measurements into account. 
 Fig. \ref{fig_magstructure} summarizes the possible magnetic structures with the propagation vector $\vec{K}$=(0,0,0) in h-LuFeO$_3$, where $\Gamma_1$, $\Gamma_2$, $\Gamma_3$ and $\Gamma_4$ are irreducible representations (IR) of the P6$_3$cm group.
 We notice that the only IR that allows for a ferromagnetic component along $c$-axis is $\Gamma_2$.\cite{Munoz2000}
 Therefore the magnetic structure at temperature $T$$<$$T_R$ has to include $\Gamma_2$, mixed with $\Gamma_1$ which  contribute to the magnetic diffraction near (100)\cite{Helton2011,supplementary,NoteGammamix}.
 For the magnetic structure at temperatures $T_R$$<$$T$$<$$T_N$, $\Gamma_1$ and $\Gamma_3$ can be ruled out since $\Gamma_1$ and $\Gamma_3$ will lead to a substantial intensity of magnetic diffraction near (100)\cite{Helton2011,supplementary}, which actually disappears at $T$$>$$T_R$.
 Therefore, the candidate magnetic structures capable of describing the magnetic structures at temperatures $T_R$$<$$T$$<$$T_N$ are $\Gamma_2$ or $\Gamma_4$.

 Assuming that the ferromagnetic moment is coming from the Dzialoshinskii-Moriya (DM) interaction,\cite{Dzyaloshinsky1958,Moriya1960,Akbashev2011,Munoz2000,Morin1950,Treves1965} the effective Hamiltonian is $H_{DM}=-\sum\limits_{i,j}{\vec{D}_{i,j}\cdot(\vec{S}_i\times\vec{S}_j)}$, where $\vec{S}_i(\vec{S}_j)$ is the spin on site $i$ ($j$), and $\vec{D}_{i,j}$ is the interaction coefficient. 
 The spin canting toward the $c$-axis allows a slight reduction of the spin-spin angles from 120 degree, lowering energy according to the DM interaction, while keeping the in-plane projections of spin-spin angles 120 degree.
 In principle, the interaction coefficients $\vec{D}_{i,j}$ do not change dramatically with temperature. 
 Since the spin structure $\Gamma_2$ generates a ferromagnetic component at $T$$<$$T_R$, the presence of $\Gamma_2$ at $T_R$$<$$T$$<$$T_N$ would also generate a similar ferromagnetic component which was not observed. 
 Therefore, the only probable spin structure is $\Gamma_4$ at $T_R$$<$$T$$<$$T_N$. 
 
 Our results in h-LuFeO$_3$ films are distinct from previous reports in the clear observations of polar structure up to 1050 K and antiferromagnetism up to 440 K, the latter of which is difficult to detect using magnetization measurements alone, especially with the strong paramagnetic background from the YSZ substrates.\cite{Bossak2004,Akbashev2011,supplementary} 
 In principle, the magnetic ordering temperature in h-LuFeO$_3$ can be affected by the following factors: 
 the exchange interaction between the Fe$^{3+}$ sites, which is supposed to be large, as indicated by the high $T_N$ (620 K) in o-LuFeO$_3$;\cite{Cheremisinoff1990} 
 the triangular spin lattice which creates the frustration and is supposed to lower the long range spin ordering temperature;\cite{Cheremisinoff1990,Klimin2003,Mitsuda,Plumer2010,Ramirez1994}
 and the magnetic anisotropy energy which affects the spin orientation and is responsible for relaxing the frustration and causing the long range magnetic order.\cite{Plumer2010,Ramirez1994} 
 Clearly, future investigations on the interplay between these factors will be needed to account for the high magnetic ordering temperature in h-LuFeO$_3$.

  In conclusion, we have demonstrated the coexistence of ferroelectricity and magnetic order at room temperature in h-LuFeO$_3$ films. 
  The polar structure and the antiferromagnetic order of the films persist up to 1050 K and 440 K respectively. 
  Besides the room temperature multiferroicity, a spin reorientation resulting in the $\Gamma_2$ magnetic structure below $T_R$=130 K is found as the origin of the net magnetic moment collinear with the electric polarization which suggests a possible linear magnetoelectric coupling.\cite{Das2013}
  The $\Gamma_2$ magnetic structure also allows for a uniaxial alignment of the non-zero R$^{3+}$ magnetic moments and a large net magnetic moment along the $c$-axis in h-RFeO$_3$; this is observed in h-YbFeO$_3$ ($\approx$3$\mu_B$/formula unit).\cite{Iida2012,Jeong2012}
  Therefore, future investigations should aim at tuning the spin reorientation up to room temperature or changing spin structure to $\Gamma_2$ right below $T_N$. This approach should render hexagonal ferrites with magnetic R$^{3+}$ the desirable multiferroic properties: large electric and magnetic polarizations as well as linear magnetoelectric couplings at room temperature.

 Research supported by the U.S. Department of Energy, Basic Energy Sciences, Materials Sciences and Engineering Division (H.M.C., H.N.L, P.C.S., T.Z.W., X.S.X.) and performed in part at the Center for Nanophase Materials Sciences (CNMS) (Z.G., N. B.) and ORNL's Shared Research Equipment (SHaRE)(M.C.) User Facility, which are sponsored at Oak Ridge National Laboratory by the Office of Basic Energy Sciences, U.S. Department of Energy. 
 We also acknowledge partial funding support from the National Basic Research Program of China (973 Program) under the grant No. 2011CB921801 (J.S.) and the US DOE Office of Basic Energy Sciences, the US DOE grant DE-SC0002136 (W.B.W.).
 Use of the Advanced Photon Source was supported by the U. S. Department of Energy, Office of Science, Office of Basic Energy Sciences, under Contract No. DE-AC02-06CH11357. 
 The work at Rutgers is supported by DOE-BES under grant No. DE-SC0008147.
 X. M. Cheng acknowledges support from the National Science Foundation under Grant No. 1053854.

$^*$ To whom correspondence should be addressed: xiaoshan.xu@gatech.edu, shenj5494@fudan.edu.cn.


\begin{references}
\bibitem{NoteRHEED} Here we use a relatively large incident angle (3.4$^\circ$) for RHEED to increase the sensitivity to the bulk part of the films.
\bibitem{NoteUnitCell} The P6$_3$cm unit cell is used for the indices of crystal directions and planes of h-LuFeO$_3$ in the rest of the paper.
\bibitem{NoteGammamix} $\Gamma_2$ can not be mixed with $\Gamma_4$ because it generates uneven spins on different Fe$^{3+}$ sites. Mixing of $\Gamma_2$ with $\Gamma_3$ is also unlikely because it involves antiphase rotations of spins on different layers of Fe$^{3+}$ sites which affects the interlayer couplings of the spins.
\end{references}
\end{document}